\documentclass{aa}
\usepackage{graphicx}

\begin{document}

\title{Metal Enrichment Processes in the Intra-Cluster Medium}

\author{S. Schindler${^1}$, W. Kapferer${^1}$, W. Domainko${^1}$,
M. Mair${^1}$, E. van Kampen${^1}$, T. Kronberger${^1}$, S. Kimeswenger${^1}$,
M.~Ruffert${^2}$, O. Mangete${^2}$, D. Breitschwerdt${^3}$}

\institute{${^1}$Institut f\"ur Astrophysik, Leopold-Franzens Universit\"at
Innsbruck, Technikerstra\ss e 25, A-6020 Innsbruck, Austria\\
http://astro.uibk.ac.at/astroneu/hydroskiteam/index.htm\\
${^2}$ School of Mathematics, University of
Edinburgh, Edinburgh EH9 3JZ, Scotland, UK\\
${^3}$ Institut f\"ur Astronomie, Universit\"at Wien, T\"urkenschanzstr. 17, 1180 Vienna, Austria}

\offprints{\email{sabine.schindler@uibk.ac.at}}

\authorrunning{S. Schindler et al.}
\titlerunning{Metal Enrichment Processes in the Intra-Cluster Medium}
\date{Received / Accepted}

\abstract{We present numerical simulations of galaxy clusters which
include interaction processes between the galaxies and the
intra-cluster gas. The considered interaction processes are galactic
winds and ram-pressure stripping, which both transfer metal-enriched
interstellar medium into the intra-cluster gas and hence increase its
metallicity. We investigate the efficiency and time evolution of the
interaction processes by simulated metallicity maps, which are
directly comparable to those obtained from X-ray observations. We find
that ram-pressure stripping is more efficient than quiet
(i.e. non-starburst
driven) galactic winds in the redshift interval between 1 and 0. The
expelled metals are not mixed immediately with the intra-cluster gas, but
inhomogeneities 
are visible in the
metallicity maps. Even stripes of higher metallicity that a single
galaxy has left behind can be seen. The spatial distribution of the
metals transported by ram-pressure stripping and by galactic winds are
very different for massive clusters: the former process yields a 
centrally concentrated
metal distribution while the latter results in an extended metal
distribution.

\keywords{Galaxies:clusters:general -- Galaxies:abundances --
Galaxies:interactions -- Galaxies:ISM --
X-ray:galaxies:clusters}
}

\maketitle


\section{Introduction}

The components of clusters of galaxies -- galaxies and intra-cluster
medium (ICM) -- interact with each other in many ways. There is more
and more observational evidence that various types of 
processes are at work, which remove interstellar medium (ISM) from the
galaxies. 
This metal enriched ISM mixes with the ICM, so that the currently
observed ICM is a 
mixture of both. Hence the metals are a good tracer for the present
and past interaction
processes  between galaxies and ICM.
The new X-ray
observations by CHANDRA 
and XMM  make it
possible to measure not only profiles but
2D distributions of metals: metallicity maps
(e.g. Schmidt et al. 2002; Furusho et al. 2003; Sanders et al. 2004;
Fukazawa et al. 2004; Hayakawa et al. 2004) showing that
the metallicity has clearly
a non-uniform, non-spherical distribution.

Already many years ago supernova-driven galactic winds were suggested
as a possible ISM transfer mechanism (De Young 1978).
Recent observations and simulations indicate
that other processes can contribute considerably, as well (e.g. Hayakawa et
al. 2004).
One process gaining increasing 
attention is ram-pressure stripping (Gunn \& Gott 1972). In the
Virgo cluster alone at least 7 ram-pressure affected
spiral galaxies have been found  (e.g. Cayatte et
al. 1990; Vollmer et al. 2004). Also other processes like galaxy-galaxy
interactions (Kapferer et al. 2005), jets from AGNs and intra-cluster supernovae (Domainko et
al. 2004) can
contribute to the metal enrichment.

Currently, several approaches are being made to explain the overall
enrichment of the ICM by taking into account galactic winds.
De Lucia et al. (2004) and Nagashima et
al. (2004) use a combination of semi-analytic
techniques and N-body simulations to calculate the overall ICM
metallicity, but they do not predict the distribution of metals in a
cluster. They find that mainly the massive galaxies contribute to the
enrichment and that there is a mild metal evolution since z=1.
Tornatore et al. (2004) do
simulations with smoothed particle hydrodynamics
that include detailed yields from type Ia and II
supernovae, but do not distinguish between the different transport
processes. They put most of their emphasis on the amount of iron produced and
on metallicity profiles.

Other groups have calculated the effect of ram-pressure stripping on
single galaxies (e.g. Abadi et al. 1999; Quilis et al. 2000; Toniazzo \&
Schindler 2001; 
Roediger \& Hensler 2004), but not the effect on the ICM.
We  take a
complementary approach by investigating the
transport of the metals from the galaxies into the intra-cluster
medium and study in detail the efficiency of the various
transport processes. In our
simulations presented here,
we have taken into account
two different transport processes --  ram-pressure stripping and
galactic winds.


\section{Simulations}

\subsection{Numerical method}

We use  N-body and hydrodynamic techniques
together with a semi-numerical phenomenological galaxy formation code
and various prescriptions for interactions between galaxies and the
intra-cluster medium. 
The core of the programme is a hydrodynamic code with shock capturing
scheme (PPM, Colella \& Woodward
1984), multiple grid refinement (Ruffert
1992) and radiative cooling. 
This hydrodynamic code is combined with an N-body tree code (Barnes \&
Hut 1986) 
-- providing actual galaxy orbits --
with constrained random fields as inital conditions (Hoffman \& Ribak 1991)
as implemented
by van de Weygaert \& Bertschinger (1996).
It also includes
an improved version of the galaxy formation code by van Kampen et
al. (1999), which  provides galaxy properties. We use a $\Lambda$CDM
cosmology ($\Omega_{\Lambda}=0.7$ ,$\Omega_{m}=0.3$, $\sigma_8=0.93$,
and $h=0.7$) with $64^3$ particles.
While the N-body code and the galaxy
formation trace the whole evolution from the beginning 
the hydrodynamic covers the
redshift interval between 1 and 0 
with time steps of about 15 Myrs
starting with the initial condition of hydrostatic equilibrium.
The hydrodynamic simulation is
calculated on 4
levels of nested grids centred at the cluster centre, the largest being
(20Mpc)$^3$ and the smallest (2.5Mpc)$^3$. The number of grid cells at
each level is $128^3$.
In addition we use various prescriptions for the different interaction
processes which transport metal enriched gas from the galaxies into
the ICM. 
These transport processes are calculated in the full
simulation volume.

\subsection{Galactic winds}

The amount of matter ejected by galactic winds is calculated with
an approach initiated by Breitschwerdt et al. (1991). The
algorithm includes thermal and cosmic ray driven winds in disk
galaxies. The wind code requires several galaxy parameters like
halo mass, disk mass, spin parameter, scale length of the
components, temperature distribution of the ISM, magnetic field
strength and gas density distributions as well as stellar density
distribution. We perform parameter studies and summarise their 
results in a look up
table for about 1000 different disk galaxies. 
The database
includes galaxies with halo masses of $1\times10^{10}$ M$_{\odot}$
- $1\times10^{12}$ M$_{\odot}$. In addition 10 different disk
masses for a given halo mass (0.01 - 0.5 $\times$ halo mass) and
10 different spin parameters for each halo-disk
combination are calculated. As
we want to investigate the contribution of galactic winds to the enrichment of
the ICM of quiet, non-starburst disk galaxies (= steady-state winds) 
the parameters of
the ISM and the star formation rate 
(1 M$_{\odot}$/yr) are kept constant. 
Following
this approach we find an average mass loss rate of about 0.3
M$_{\odot}$/yr. 
In case of a mass
loss due to a galactic winds  the ISM is transferred
with the corresponding
metallicity and a temperature of $5\times10^{6}$~$\mathrm{K}$ 
into the hydrodynamic simulation at the position of the galaxy.

\subsection{Ram-pressure stripping}

Galaxies in galaxy clusters can suffer mass loss from interaction with
 the ICM, e.g. ram pressure acts on the gas disk of galaxies moving
 through the ICM. We assume that gas is stripped off beyond the
radius of the galaxy where the restoring gravitational force is
equal to the force due to ram-pressure (stripping radius, Gunn \&
Gott 1972). We assume for disk galaxies an exponential stellar
disk and gas disk. The validity of this approach has been shown by
simulations and by comparisons
of galaxies in the Virgo
 cluster affected by ram pressure (Abadi et al. 1999).
For supersonic galaxies we follow the Rankine Hugoniot condition to derive
the properties of the galaxies surrounding ICM. Mass loss of galaxies moving
inclined through the ICM is scaled with the cosine of the inclination
angle with respect to the mass loss derived for an uninclined galaxy.
Galaxies which are stripped are assumed to have a
truncated gas disk and they only lose more gas if
increased ram pressure further reduces the stripping radius.


\section{Results}
\label{results.sec}

We present simulations of clusters with different mass: a massive
($1.3\times10^{15}$ M$_{\odot}$, Cluster 1)
and a less massive cluster ($7.4\times10^{14}$
M$_{\odot}$, Cluster 2). Cluster 2 undergoes
a major merger (mass ratio 1:3), whereas Cluster 1 
only has small merger events. 

In Fig.~1 the X-ray emission weighted metallicity maps are presented
for Clusters 1 and 2: only the enrichment that has taken place since
redshift 1 has been taken into account. 
Cluster 1 is shown at redshift $z=0$ with
two different enrichment mechanisms, ram-pressure
stripping (Panel~1a) and  galactic winds (Panel~1b). 
The two mechanisms result in a very different spatial distribution. 
Ram-pressure stripping is very efficient in the centre, because 
there the ICM density is very high  ($\sim6\times10^{-27}$
$\mathrm{g/cm^{3}}$ within a radius of 0.1 $\mathrm{Mpc}$) and leads to
a centrally concentrated metallicity distribution. On the
other hand the high central ICM density yields a high pressure which 
suppresses galactic winds
resulting in a very extended distribution.
Apart from the differences in the centre both processes
yield pronounced inhomogeneities in the abundances.
Single galaxies leave traces of metals in shapes of stripes 
behind them. 
Both the centrally concentrated metallicities and the
general inhomogeneities are in good agreement with observations. 
The
increase of metallicity (X-ray emission weighted) 
since redshift 1 averaged over
the central cluster region within a radius of 400 kpc is 0.09 and
$4\times10^{-8}$ in solar units
for the ram-pressure and the winds model, respectively.

The situation looks very different in a less massive cluster. In
Cluster 2 the winds are less suppressed. Therefore the winds contribute
also to the central metallicity.  
Inhomogeneities in the abundances do not disperse immediately 
(compare insets) even
though the ICM in this cluster is stirred by a merger.
The
stripes are gradually spread out and at the end a roughly homogeneous
region of high metallicity is present at the centre. A given
inhomogeneity at t=1.7 Gyrs will be spread out on
average over a 30 times larger volume at the end of the simulation
(t=8 Gyrs),  depending strongly  on the local dynamics of the
surrounding ICM, e.g. shocks. 

Non-starburst driven winds are less efficient in transporting gas than
ram-pressure stripping in the time interval between redshift 1 and 0
investigated here. 
During the non-merger phases of Cluster 2
ram-pressure stripping is about a factor of 5 more efficient than
non-starburst driven winds.
Figure 2 shows the evolution of the metal mass loss for ram-pressure
stripping and quiet galactic winds in a merger cluster. The different
efficiencies are clearly visible.
The
increase of metallicity  (X-ray emission weighted) 
since redshift 1 averaged over
the central cluster region within a radius of 400 kpc is 0.07 and 0.0035
in solar units
for the ram-pressure and the winds model, respectively.

The mass loss due to ram-pressure stripping shows large fluctuations
compared to  galactic winds, because
non-starburst driven winds have an
almost constant mass loss over a long period of up to 5 Gyrs, whereas
ram-pressure stripping removes huge amounts of gas from a single
galaxy on much shorter time scales. During the two merger events (at
3 Gyrs and 5.5 Gyrs) the mass loss due to ram pressure is much higher
because of the enhanced ICM pressure. In contrast 
non-starburst driven galactic winds are suppressed in 
galaxies located
in the high ICM pressure region.
Therefore the mass loss due to winds is slightly lowered during mergers.

\begin{figure*}
\centering
\includegraphics[width=\textwidth]{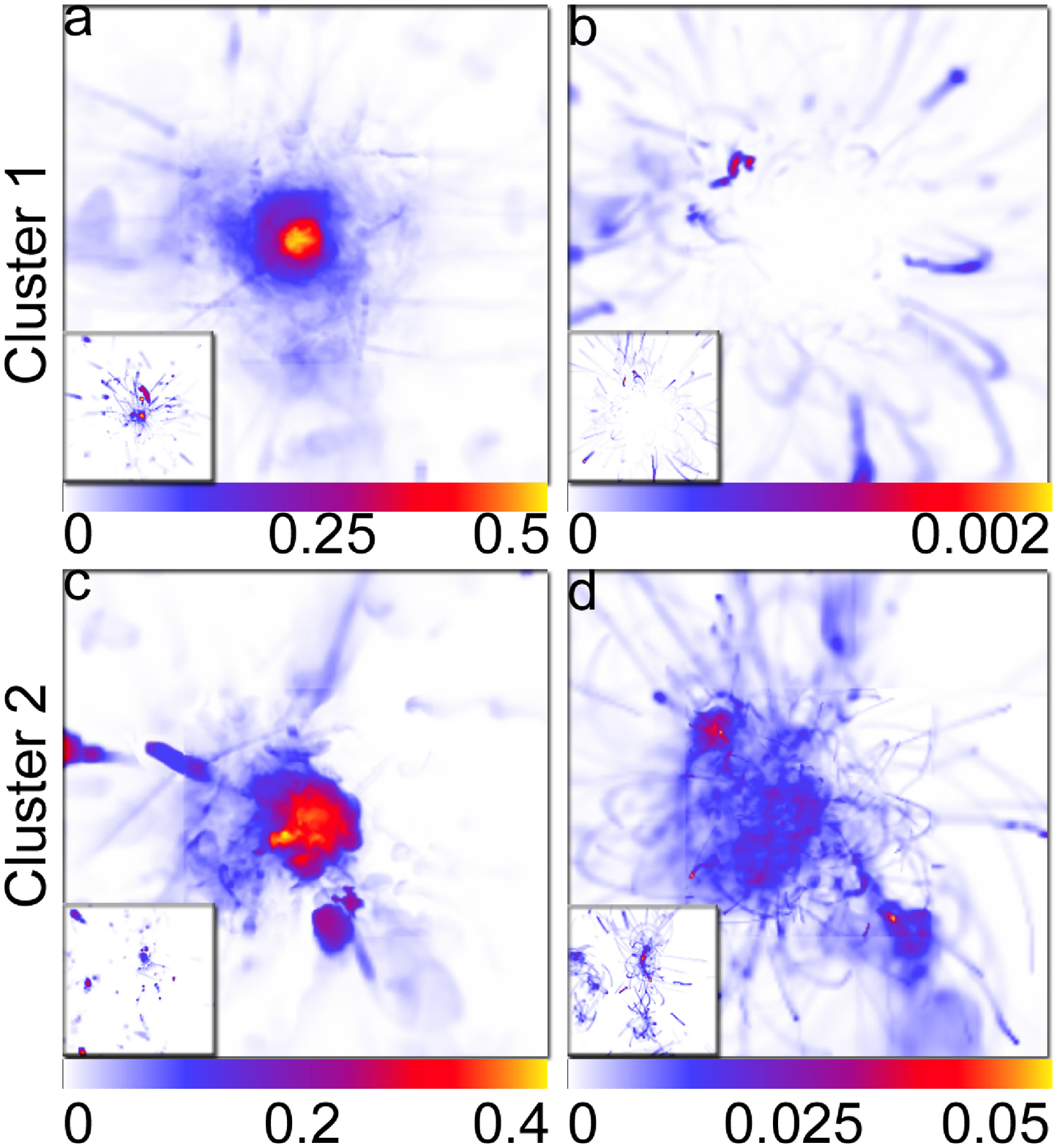}
\caption{X-ray weighted metal maps of two simulations at present epoch
and insets at an epoch 6.7~Gyrs ago (a and b, Cluster 1:
a massive galaxy cluster; c and d, Cluster 2: a merging
cluster). Panels a
and c show the metal maps for ram-pressure stripping and panels
b and d those for galactic winds (non-starbursts). The quantities are
given in solar abundances and the size of the maps is 5
$\mathrm{Mpc}$ on the side. Ram-pressure stripping always yields a
centrally concentrated metallicity distribution. The distribution of
metals transported by winds depends strongly on the cluster mass: in
a massive cluster (b) the winds are suppressed in the centre resulting
in a low abundance there. The comparison with the insets shows how the
inhomogeneities spread/evolve.
}
\label{image}
\end{figure*}

\begin{figure}[h]
\includegraphics[width=8.8cm]{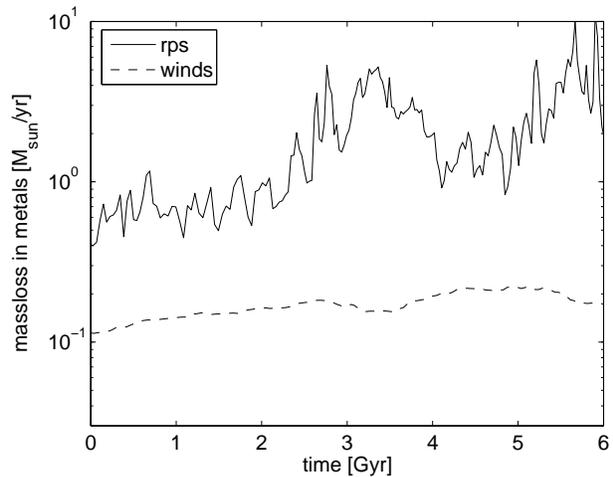}
\caption{Evolution of the metal transfer from the ISM to the ICM 
in solar masses per year
for model Cluster 2. The metal mass loss for ram-pressure stripping and for
non-starburst driven galactic winds are shown. } \label{plot}
\end{figure}


\section{Summary and Conclusions}

Different enrichment mechanisms -- ram-pressure stripping and galactic
winds -- yield different metallicity
distributions.

\begin{itemize}
\item{}In massive clusters ram-pressure
stripping provides a much more centrally concentrated distribution than
galactic winds, because galactic winds can be suppressed in the
cluster centre while ram-pressure stripping is most efficient there.

\item{}Both processes yield abundance distributions with 
lots of inhomogeneities in good agreement with recent X-ray
observations. 

\item{}The inhomogeneities are not dispersed immediately, but
are gradually spread out. 

\item{}The efficiency of gas transfer due to ram-pressure stripping is
higher compared to non-starburst driven galactic winds in the redshift
interval between 1 and 0.
\end{itemize}

The comparison of these results with new detailed observed
metallicity maps will yield insights into the various interaction processes
between the ICM and the cluster galaxies. The knowledge on their efficiency and
timescales will provide information on the formation and evolution  of
galaxies and their hosting clusters. In the future we will extend
the simulations to higher redshifts ($z>1$) 
so that we will be able to compare the
simulated metallicities directly with the observed metallicity maps.
Further simulations including not
only more interaction processes, 
but also distinguishing between different
chemical elements are planned.


\begin{acknowledgements}
This work was supported by the Austrian Science Foundation FWF
(P15868), UniInfrastruktur 2004, Tiroler Wissenschaftsfonds,  
AUSTRIAN GRID and by a University of
Innsbruck scholarship. Edmund Bertschinger and Rien van de Weygaert are acknowledged
for providing their constrained random field code.
\end{acknowledgements}


\end{document}